\documentclass[twocolumn,11pt]{wseas}

\usepackage{abstract, fancyhdr,rotating}
\usepackage{color,graphicx,amssymb,amsbsy,amsmath}
\usepackage{times}
\renewcommand{\thefootnote}{\fnsymbol{footnote}}

\title{Spatial correlations in SIR epidemic models}
\author{HENRYK FUK\'S\footnotemark[2] 
\and RYAN DUCHESNE\footnotemark[2]
\and ANNA T. LAWNICZAK\footnotemark[4]
\vspace{1mm} \and 
  Department of Mathematics \\
Brock University\footnotemark[2] \\
St. Catharines, Ontario L2S 3A1 \\ 
CANADA \vspace{2mm} \\
  Department of Mathematics and Statistics\\
University of Guelph\footnotemark[4]\\
Guelph, Ontario N1G 2W1\\ CANADA 
\vspace{2mm}\\
Corresponding author's email: hfuks@brocku.ca}

\begin{document}
\twocolumn[
\vspace*{-1cm}
  \maketitle
  \vspace{-18mm}
\begin{onecolabstract}
  \noindent \emph{Abstract: - } We investigate the role of  global
  mixing in epidemic processes. We first construct a simplified model
  of the SIR epidemic using a realistic population distribution.  Using
  this model, we examine possible mechanisms for destruction of
  spatial correlations, in an attempt to produce correlation curves
  similar to those reported recently for real epidemiological data. We
  find that introduction of a long-range interaction destroys spatial
  correlations very easily if neighbourhood sizes are homogeneous. For
  inhomogeneous neighbourhoods, very strong long-range coupling is
  required to achieve a similar effect.
\end{onecolabstract}
\vspace{3mm}
]


\renewcommand{\thefootnote}{\alph{footnote}}

\addtolength{\abovedisplayskip}{-1mm}
\addtolength{\belowdisplayskip}{-1mm}
\addtolength{\belowcaptionskip}{-0.5mm}

\section{Introduction}
The main mechanism of transmission of infectious diseases is usually
the direct contact of susceptible individuals with an infective one.
Since this contact is normally highly localized in space, it is quite
natural to expect that space should play an important role in dynamics
of infectious diseases. There is a clear evidence that some infectious
diseases in animal populations spread geographically. A well-known
example is the spatial advance of fox rabies in Europe, which seems to
have started in Poland in 1939, and has moved steadily westward at a
rate of 30-60 km per year \cite{Murray2003b}. Similar patterns of
spread have been observed in the epizootic of rabies among raccoons in
eastern United States and Canada. It started in 1977 in an area on the
West Virginia-Virginia border and has moved at a rate of 30-40 km per
year \cite{rabies2000}.

In human populations, the spread of the Black Death in Europe from
1347--1350 is the most often quoted example \cite{Langer1964}.
Introduced to Italy in 1347, it spread up through Europe at 300-500 km
per year. For other diseases the spatial effects appear to be somewhat
less pronounced. For example, even though in the past influenza
pandemics used to reveal spatial patterns \cite{Rvachev95}, there is a
very strong evidence that in recent times the spread of influenza is
statistically uniform in space.  Bonabeau et al. \cite{Bonabeau1998}
examined the spatial correlation structure of the influenza epidemic
for the epidemic of winter 1994--5, using high-quality data collected
by a large network of general practitioners in France. They found that
at least for  influenza epidemics, space does not play any
important role, and the spread of the disease is dominated by the
mean-field dynamics.

The goal of this paper is to investigate the role of  global mixing
in the spread of epidemics. We first construct a simplified model
of the SIR epidemic based on realistic population distribution.
Using this model, we investigate possible mechanisms for
destruction of spatial correlations, trying to achieve similar effects
as those reported by Bonabeau et al. \cite{Bonabeau1998}.

\section{Description of the model}
In order to study the influence of  global mixing on the spread of
epidemics we use a model based on interacting particle systems, in the
spirit of our earlier work \cite{paper15,paper18}.  Models of this
type take various forms, ranging from stochastic interacting particle
models \cite{durret94} to models based on cellular automata or coupled
map lattices~\cite{schon93,bc93,duryea99,benyo2000}.

Consider a set of $N$ individuals, labelled with consecutive integers
$1,2\ldots, N$. This set of labels will be denoted by $\mathcal{L}$.
We assume that each individual can be in three distinct states,
susceptible (S), infected (I) or removed (R).  There are two ways to
change the state of a single individual.  A susceptible individual who
comes in direct contact with an infected individual can become
infected with probability $p$.  Infected individual can become removed
with probability $q$. The precise description of the model is as
follows.

The state of the $i$-th individual at the time step $k$ will be
described by a Boolean vector variable $ \boldsymbol {\eta}(i,k)=
\langle\eta_S(i,k),\eta_I(i,k),\eta_R(i,k)\rangle$, where
$\eta_\tau(i,t)=1$ if the $i$-th individual is in the state $\tau$,
where $\tau \in \{S,I,R\}$, and $\eta_\tau(i,k)=0$ otherwise. We
assume that $i=1,2,\ldots, N$ and $k\in\mathbb{N}$, i.e., the time is
discrete. Hence, the vector $ \boldsymbol {\eta}(i,k)$ can be in one
of the following states: $ \boldsymbol {\eta}(i,k)= \langle 1,0,0
\rangle$ for a susceptible individual, $ \boldsymbol
{\eta}(i,k)=\langle 0,1,0 \rangle$ for an infected individual, and $
\boldsymbol {\eta}(i,k)=\langle 0,0,1 \rangle$ for a removed
individual. No other values of $\boldsymbol {\eta}(i,k)$ are possible
in SIR epidemic model. In SIR model an individual can only be at one state at any give time and transitions occur only from susceptible to infected and from infected to removed. The removed does not become susceptible or infected again in SIR model. Therefore, SIR model is suitable for studying spread of influenza in the same season because the same type of influenza virus can infect an individual only once and once the individual is recovered from the flu it becomes immune to this type of virus.

We further assume that at the time step $k$ the $i$-th individual can
interact with individuals from a subset of $\mathcal{L}$, to be
denoted by $C(i,k)$. Using this notation, we obtain
\begin{eqnarray}\label{sirdyn1}
  \eta_S(i,k+1)&=&\eta_S(i,k) \prod_{j \in C(i,k)}
  \overline{X_{i,j,k}\eta_I(j,k)},\\\label{sirdyn2}
  \eta_I(i,k+1)&=& \eta_S(i,k) \Big(1-\prod_{j \in C(i,k)}
  \overline{X_{i,j,k}\eta_I(j,k)}\Big)\nonumber 
  \\ &&+ \eta_I(i,k) \overline{Y}_{i},\\\label{sirdyn3}
  \eta_R(i,k+1)&=&  \eta_R(i,k) +    \eta_I(i,k) Y_{i},
\end{eqnarray}
where $X_{i,j,k}$ is a set of iid Boolean random variables such that
$Pr(X_{i,j,k}=1)=p$, $Pr(X_{i,j,k}=0)=1-p$, and $Y_{i}$ is a set of
iid Boolean variables such that $Pr(Y_{i}=1)=q$, $Pr(Y_{i}=0)=1-q$.

Note that $X_{i,j,k}=1$ means that the disease has been transmitted
from the $j$-th individual to the $i$-th individual at  time step
$k$. If at least one of the random variables $X_{i,j,k}$ in the
product $\prod_{j \in C(i,k)} \overline{X_{i,j,k}\eta_I(j,k)}$ takes
the value 1, then the product becomes 0, and we obtain
$\eta_S(i,k+1)=0$, meaning that the $i$-th individual changes its
state from susceptible to infected.

The crucial feature of this model is the set $C(i,k)$, representing all
individuals with whom the $i$-th individual may have interacted at the
time step $k$. In a large human population, it is almost impossible to
know $C(i,k)$ for each individual, so we make some simplifying
assumptions. First of all, it is clear that the spatial distribution
of individuals must be reflected in the structure of $C(i,k)$. We have
decided to use realistic population distribution for Southern and
Central Ontario using census data obtained from Statistic Canada
\cite{statcan1,statcan2}. The selected region is mostly surrounded by
waters of Great Lakes, forming natural boundary conditions.  The data
set specifies population of so called ``dissemination areas'' , i.e.,
small areas composed of one or more neighbouring street blocks. We had
access to longitude and latitude data with accuracy of roughly
$0.01^\circ$, hence some dissemination areas in densely populated
regions had the same geographical coordinates. We combined these
dissemination areas into larger units, to be called ``modified
dissemination areas'' (MDA).

We will now define the set $C(i,k)$ using the concept of MDAs. This
set will be characterized by two positive integers $n_c$ and $n_f$.
Let us label all MDAs in the region we are considering by integers $m
=1,2, \ldots ,M$, where in our case $M=5069$. For an individual $i$
belonging to the $m$-th MDA, the set $C(i,k)$ consists of all
individuals belonging to the $m$-th MDA, plus all individuals
belonging to $n_c$ MDAs nearest to $m$, plus $n_f$ MDAs randomly
selected among all remaining MDAs.  While the ``close neighbours'',
i.e., $n_c$ nearest MDAs, will not change with time, the ``far
neighbours'', i.e., $n_f$ randomly selected MDAs, will be randomly
reselected at each time step.

\section{Mean Field}
The model described in the previous section involves strong spatial
coupling between individuals. Before we describe consequences of this
fact, we will first construct a set of equations which approximate
dynamics of the model under the assumption of ``perfect mixing'',
i.e., neglecting the spatial coupling.

The state of the system described by eq.
(\ref{sirdyn1}--\ref{sirdyn3}) at time step $k$ is determined by
the states of all individuals and is described by the Boolean random
field $\boldsymbol{\eta}(k)=\{\boldsymbol{\eta}(i,k): i=0,\ldots,N\}$.
The Boolean field $\{\boldsymbol{\eta}(k): i=0,1,2\ldots\}$ is then a
Markov stochastic process.

By taking the expectation $E_{\boldsymbol{\eta}(0)}$ of this Markov
process when the initial configuration is $\boldsymbol{\eta}(0)$, i.e.
$\rho_{\tau}(i,k)=E_{\boldsymbol{\eta}(0)}\left[\eta_{\tau}(i,k)
\right]$ for $\tau \in \{S,I,R\}$, we get the probabilities of the
$i$-th individual being susceptible, infected or removed at time $k$.

In the mean field approximation, we assume independence of random
variables $\eta_{\tau}(i,k)$. Hence, the expected value of a product of
such variables is equal to the product of expected values. Under this
assumption, taking expected values of both sides of equations
(\ref{sirdyn1}--\ref{sirdyn3}) we obtain
\begin{eqnarray}\label{sirmf1}
 \rho_S(i,k+1)&=&\rho_S(i,k) \prod_{j \in C(i,k)}
 (1-p\rho_I(j,k)),\\\label{sirmf2}
  \rho_I(i,k+1)&=& 
  \rho_S(i,k) \Big(1-\prod_{j \in C(i,k)}
 (1-p\rho_I(j,k))\Big) \nonumber \\
 &&+ \rho_I(i,k) (1-q),\\\label{sirmf3}
 \rho_R(i,k+1)&=&  \rho_R(i,k) +    \rho_I(i,k) q.
\end{eqnarray}
Since mean field approximations neglect spatial correlations, we
further assume that $\rho_{\tau}(i,k)$ is independent of $i$, i.e.,
$\rho_{\tau}(i,k)=\rho_{\tau}(k)$. Even though sets $C(i,k)$ have
different number of elements for different $i$ and $k$, for the
purpose of this approximate derivation we assume that they all have
the same number of elements $(1+n_c+n_f)D$, where $D$ is the average
MDA population. All these assumptions lead to
\begin{align}\label{mf1}
  \rho_S(k+1)&=\rho_S(k) (1-p\rho_I(k))^{(1+n_c+n_f)D},\\\label{mf2}
  \rho_I(k+1)&= \rho_I(k) + \rho_S(k) \nonumber\\
  -\rho_S&(k)(1-p\rho_I(k))^{(1+n_c+n_f)D} -q \rho_I(k) ,\\\label{mf3}
  \rho_R(k+1)&= \rho_R(k) + q \rho_I(k).
\end{align}
The third equation in the above set is obviously redundant, since
$\rho_S(k)+\rho_I(k)+\rho_R(k)=1$.

Similarly to the classical Kermack-McKendrick model, mean field
equations (\ref{mf1})-(\ref{mf3}) exhibit a threshold phenomenon.
Depending on the choice of parameters, we can have
$\rho_I(k)<\rho_I(0)$ for all $k$, meaning that the infection dies
out, or we can have an outbreak of the epidemic, meaning that
$\rho_I(k)>\rho_I(0)$ for some $k$. The intermediate scenario of
constant $\rho_I(k)$ will occur when $\rho_I(k)=\rho_I(0)$, i.e., when
\begin{equation}\label{trecond}
     \rho_S(0)
 -\rho_S(0)(1-p\rho_I(0))^{(1+n_c+n_f)D} -q  \rho_I(0)=0.
\end{equation}
Assuming that initially there are no individuals in the removed group,
we have $\rho_S(0)=1-\rho_I(0)$. Furthermore, if $(1+n_c+n_f)D$ is
large, we can assume $(1-p\rho_I(0))^{(1+n_c+n_f)D}\approx
1-p(1+n_c+n_f)D\rho_I(0)$.  Solving eq. (\ref{trecond}) for $q$ under
these assumptions we obtain
\begin{equation}\label{separatMF}
 q=\Big(1-\rho_I(0)\Big)(1+n_c+n_f)D p.
\end{equation}
Thus, assuming the mean field approximation the epidemic can occur
only if $q<\Big(1-\rho_I(0)\Big)(1+n_c+n_f)D p$.

\section{Dynamics of the model}
The mean-field equations derived in the previous section depend only
on the sum of $n_c$ and $n_f$. This means, for example, that the model
with $n_c=5$, $n_f=0$ and the model with $n_c=0$, $n_f=5$ will have
the same mean field equations. However, the actual dynamics of these
two models will be very different. Depending on the relative size of
$n_f$ and $n_c$, the epidemic may propagate or die out, as we will see
in the following analysis.

\begin{figure}
  \centering
  \includegraphics[scale=0.7]{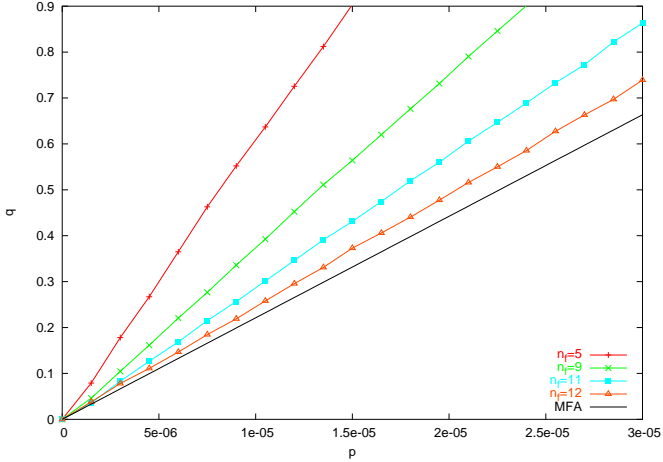}
  \caption{Graphs of critical lines for $n_f=5,9,11$, and $12$. In all
    cases, $n_c=12-n_f$. Solid line represents mean field
    approximation.}
   \label{phasetran1}
\end{figure}

Let  $N_\tau(k)$ be the expected value of the total number of
individuals belonging to class $\tau \in \{S,I,R\}$,
$$
N_\tau(k)=E_{\boldsymbol{\eta}(0)}\left( \sum_{i=1}^N \eta_\tau(i,k)
  )\right).
$$
We say that an epidemic occurs if there exists $k>0$ such that
$N_I(k)>N_I(0)$.  For fixed $p$, $n_f$ and $n_c$, there exists a
threshold value of $q$ to be denoted by $q_c$, such that for each
$q<q_c$ an epidemic occurs, and for $q>q_c$ it does not occur.
Obviously $q_c$ depends on $p$, and this is illustrated in
Figure~\ref{phasetran1}, which shows graphs of $q_c$ as a function of
$p$ for several different values of $n_f$ and $n_c$, where
$n_f+n_c=12$. The graphs were obtained numerically by direct computer
simulations of the model.  The condition $n_f+n_c=12$ means that the
size of the neighbourhood is kept constant, but the proportion of
``far neighbours'' (represented by $n_f$) to ``close neighbours''
(represented by $n_c$) varies.
Figure~\ref{phasetran1} also shows the mean-field line given
by eq. (\ref{separatMF}).

\section{Spatial correlations}
\begin{figure}[t]
   \centering
   \includegraphics[scale=0.43]{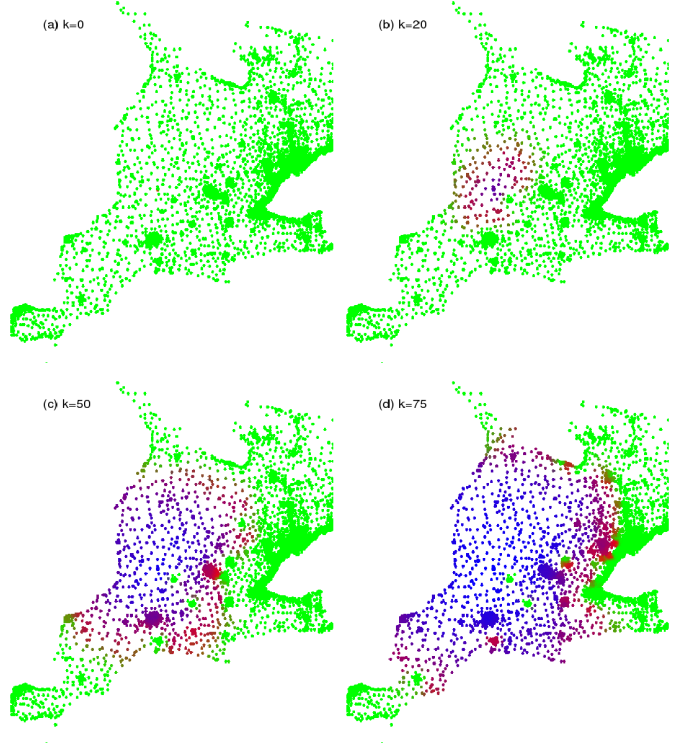}
   \caption{Example of a propagating epidemic front for $n_c=12$,
     $n_f=0$, $p=0.00005$, $q=0.05$, with (a) $k=0$, (b) $k=25$, (c)
     $k=50$ and (d) $k=75$. Modified dissemination areas are
     represented by pixels colored according to density of individuals
     of a given type, such that the red component represents density
     if infected, green density of susceptibles, and blue density of
     removed individuals.}
   \label{fig:front}
\end{figure}

As demonstrated in the previous section, the relative size of
parameters $n_f$ and $n_c$ controls dynamics of the epidemic process
in a significant way, shifting the critical line up or down. When
$n_c=0$, i.e., when there are no ``far neighbours'', the epidemic
process has a strictly local nature, and we can observe well defined
epidemic fronts propagating in space. This is illustrated in Figure
\ref{fig:front}, where the epidemic starts at $k=0$ at a single
centrally located MDA (Figure \ref{fig:front}a), with $n_c=12$,
$n_f=0$. Modified dissemination areas are represented by pixels
colored according to the density of individuals of a given type, such
that the red component of the color represents density of infected
individuals, green density of susceptibles, and blue density of
removed individuals. By density we mean the number of individuals of a
given type divided by the population of the MDA.
Epidemic wave propagating outwards can be clearly seen in subsequent
snapshots (c), (d) and (e). The front is mostly red, meaning that the
bulk of infected individuals is located at the front. After these
individuals gradually recover, the center becomes blue.

Let us now consider slightly modified parameters, taking $n_c=11$,
$n_f=1$. This means that we now replace one ``close'' MDA by one
``far'' MDA. This does not seem to be a significant change, yet the
effect of this change is quite spectacular.
\begin{figure}[t]
   \centering
   \includegraphics[scale=0.43]{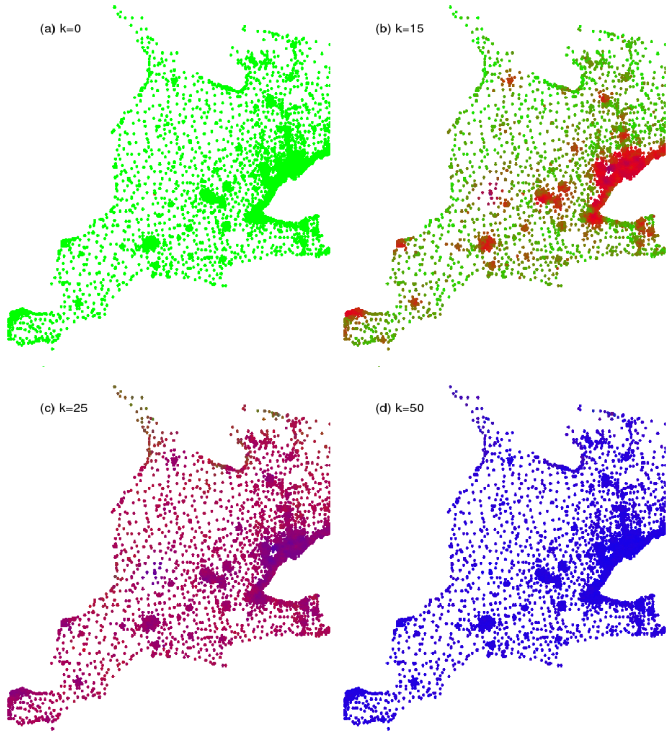}
   \caption{Development of the epidemic for $n_c=11$, $n_f=1$,
     $p=0.00005$, $q=0.05$, with (a) $k=0$, (b) $k=15$, (c) $k=25$ and
     (d) $k=50$. Colour coding is the same as in the previous figure.}
   \label{fig:frontdest}
\end{figure}
As we can see in Figure~\ref{fig:frontdest}, the epidemic propagates
much faster, and there are no visible fronts.  The disease quickly
spreads over the entire region, and large metropolitan areas become
red in a short time, as shown in Figure~\ref{fig:frontdest}(b).  This
suggests that infected individuals are more likely to be found in
densely populated regions, and their distribution is dictated by the
population distribution -- unlike in Figure~\ref{fig:front}, where
infected individuals are to be found mainly at the propagating front.

In order to quantify this observation, we use a spatial correlation
function for densities of infected individuals defined as
$$ h(r,k)=\left\langle \eta_I(i,k) \eta_I(j,k) \right\rangle_{r\leq
d(i,j) \leq r+\Delta r},
$$
where $d(i,j)$ is the distance between $i$-th and $j$-th individual,
and $<\cdot >$ represents averaging over all pairs $i$, $j$ satisfying
condition $r\leq d(i,j) \leq r+\Delta r$. In subsequent considerations
we will take $\Delta r=1\, \mathrm{km}$. The distance between two
individuals is defined as the distance between MDAs to which they
belong.

\begin{figure}
  \begin{center}
    \includegraphics[scale=0.9]{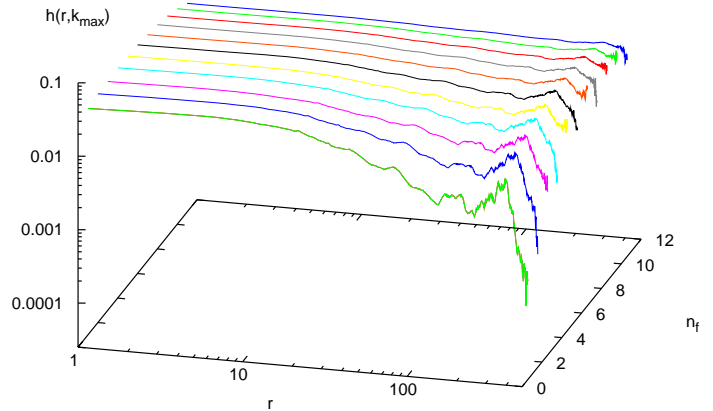}
  \end{center}
  \caption{Graphs of the correlation function $h(r,k_{max})$ for
    different values of $n_f$, where $p=0.000015$, $q=0.2$, and
    $n_c+n_f=12$.}\label{cor3d}
\end{figure}

Consider now a specific example of the epidemic process described by
eq. (\ref{sirdyn1}-\ref{sirdyn3}), where $p=0.000015$, $q=0.2$, and
$n_c+n_f=12$. For this choice of parameters epidemics always occur as
long as $n_f>0$.  Figure \ref{cor3d} shows graphs of the correlation
functions $h(r,k_{max})$ at the peak of each epidemic, so that
$k_{max}$ is the time step at which the number of infected individuals
achieves its maximum value.
An interesting phenomenon can be observed in that figure: while the
increase of the proportion of ``far'' neighbours does destroy spatial
correlations, one needs very high proportion of ``far''neighbours to
make the correlation curve completely flat. In \cite{Bonabeau1998} it
is reported that for influenza epidemics $h(r,k_{max})\sim r^{0.04 \pm
  0.03}$.  If we fit $h(r,k_{max})=C r^\alpha$ curve to the
correlation data shown in Figure \ref{cor3d}a, we obtain values of the
exponent $\alpha$ as shown in Table 1.  In order to obtain $\alpha$ of
comparably small magnitude as reported in \cite{Bonabeau1998}, one
would have to take $n_f$ equal to at least $10$, meaning that $77 \%$
of neighbours would have to be ``far neighbours''.
\begin{table}[t]
\begin{center}
 \begin{tabular}{|c|c|c|}
   \hline $n_c$ & $n_f$ & $\alpha$ \\ \hline
   \hline 11 & 1 & -0.72        +/- 0.03  \\
   \hline 10 & 2 & -0.45        +/- 0.02 \\
   \hline 9 & 3  & -0.32        +/- 0.01 \\
   \hline 8 & 4  & -0.27        +/- 0.01 \\
   \hline 7 & 5  & -0.191        +/- 0.007 \\
   \hline 6 & 6  & -0.179           +/- 0.009 \\
   \hline 5 & 7  & -0.120        +/- 0.005 \\
   \hline 4 & 8  & -0.115        +/- 0.006 \\
   \hline 3 & 9  & -0.071       +/- 0.003 \\
   \hline 2 & 10 & -0.057       +/- 0.002 \\
   \hline 1 & 11 & -0.047       +/- 0.003 \\
   \hline
\end{tabular}
\end{center}
\caption{Values of the exponent $\alpha$
  obtained by fitting $h(r,k_{max})=C r^\alpha$  to simulation data.}
\end{table}
In reality, this would require that $77\%$ of all individuals one
interacted with were not his/her neighbours, coworkers, etc., but
individuals from randomly selected and possibly remote geographical
regions.  This is clearly at odds with our intuition regarding social
interactions, especially outside large metropolitan areas. This
prompted us to investigate further and to find out what is responsible
for this effect.

Upon closer examination of spatial patterns generated in simulations
of our model, we reached the conclusion that the inhomogeneity of
population sizes in neighbourhoods $C(i,k)$ makes spatial correlations
so persistent.  Since different MDAs have different population sizes,
we expect that some individuals will have larger neighbourhood
populations than others, and as a result they will be more likely to
get infected, even if the proportion of infected individuals is the
same in all MDAs.  This will build up clusters of infected individuals
around populous MDAs.

To test if this is indeed the factor responsible for strong spatial
correlations in our model, we replaced all MDA population sizes with
constant population size $D$, i.e., average MDA population size. As
expected, graphs of the correlation functions obtained in this case
were are all essentially flat, with the exponent $\alpha$ close to
zero even in the case of $n_f=1$, when we obtained $\alpha=0.023 \pm
0.002$.

\section{Conclusions}
As demonstrated in the previous section, spatial correlations are
difficult to destroy if neighbourhood sizes are inhomogeneous.  Very
strong mixing, i.e., very significant amount of long-range
interactions is required to obtain flat correlations curves.  On the
other hand, for homogeneous neighbourhood sizes, even relatively small
long-range interaction immediately forces the process into the
perfect-mixing regime, resulting in the lack of spatial correlations.

There is a strong evidence that in recent times epidemics of influenza
do not produce significant spatial correlations \cite{Bonabeau1998} in
spite of the heterogeneity of the population distribution. One can
speculate that this must be due to one of the following two factors:
either most of our daily interactons are long-ranged, or conversely,
most interactions are short-ranged, but the number of interactions per
unit time does not vary too greatly from individual to individual.

We suspect that the answer depends on the social and economic
structure of the underlying community or adjacent communities. The
first scenario applies to large metropolitan areas and conurbations
with a large number of commuters, while the second scenario is more
likely for small communities without much interaction with the outside
 word.
The above considerations also indicate that a more realistic model will have to
separate two aspects of the interaction. First of all, with how many
individuals does a given person interact with per unit time, e.g., per
day, and how is this number distributed?  This could conceivably be
determined experimentally without much difficulty. The second aspect
is much harder, though: from what pool are these individuals chosen
and how?  While an accurate answer to this question does not seem to
be possible, we hope to get at least some insight from analysis of
travel and traffic data. This work is currently in progress and will
be presented elsewhere.

{\small
\paragraph*{Acknowledgement} 
Henryk Fuk\'s and Anna T. Lawniczak acknowledge partial financial
support from the Natural Science and Engineering Research Council
(NSERC) of Canada.

\vskip 1.5cm

\bibliographystyle{wseas}

}

\end{document}